\documentclass[a4paper,12pt]{article}
\usepackage{amsmath}
\usepackage{amsfonts}
\usepackage{amssymb}
\usepackage{graphicx}
\usepackage{multirow}
\usepackage{slashed}
\usepackage[pdftex]{hyperref}                 
\hypersetup{                            
             colorlinks = true,
	     linkcolor = blue,
	     linktocpage = true,
             citecolor = blue,
             urlcolor = blue
           }
\def\hhref#1{\href{http://arxiv.org/abs/#1}{#1}}

\def\be{\begin{equation}}
\def\ee{\end{equation}}

\def\bea{\begin{eqnarray}}
\def\eea{\end{eqnarray}}

\def\np2{\frac{\eta p^2}{M^2}}
\def\E1{E_e^\prime}
\def\p1{p^\prime}
\def\sw{\sin^2\theta_W}
\def\sws{\sin^4\theta_W}
\def\nk4{\frac{\eta k^4}{M^2}}
\def\ta{\tilde{\theta}}
\def\tb{\tilde{\theta_1}}

\begin{document}

\begin{titlepage}

\rightline{PRL-TH/AP-11/2}
\vskip 2.5cm
\begin{center}
{\Large {\bf Neutrino processes with power law dispersion relations}}
\vskip 2cm
{\large Subhendra Mohanty and Soumya Rao}\\
\vskip 0.5cm
{\it Physical Research Laboratory, Ahmedabad 380009,
India}\\
\vskip 1cm

\begin{abstract}

We compute various processes  involving neutrinos in the initial and/or final state and we
assume that neutrinos have energy momentum relation with a general  power law $E^2 =p^2+ \xi_n p^n$
correction due to Lorentz invariance violation.  We find that for $n>2$ the bounds on $\xi_n$ from direct
time of flight measurement are much more stringent than from constraining the neutrino Cerenkov decay process.

\end{abstract}
\end{center}
\end{titlepage}

\section{Introduction}

The OPERA observation \cite{Opera} of superluminal neutrinos has been ruled out by the measurement by 
ICARUS \cite{Icarus12} which puts an upper bound 
on the superluminality of neutrinos at $\delta < 2.3 \times 10^{-7}$.  
 An phenomenological consequence of neutrinos being superluminal 
was the observation by Cohen and Glashow (GC)\cite{GC} and others
\cite{Bi,Maccione,Mattingly,Carmona} that superluminal neutrinos are kinematically allowed
to emit pairs of $e^+ e^-$ and would thereby lose most of their energy during the
CERN-Gran Sasso flight of 730 km \cite{GC}. ICARUS experiment \cite{Icarus} searched for the
Cerenkov emission of electrons in the same CERN-CNGS to Gran Sasso neutrino beams, where
according to the GC calculation, $63\%$ of the neutrinos are expected to decay and no
anomalous  $e^+ e^-$ events were detected. Using this a much stronger bound $\delta < 2.5 \times 10^{-8}$
can be put on the superluminal neutrinos. Another problem with superluminal neutrinos
which has been discussed \cite{Mestres, Bi, Cowsik,Altschul} is that the change in
neutrino energy-momentum relation restricts the phase-space for the  $\pi\rightarrow \nu
\mu$ process and the pion lifetime would be larger 
for the OPERA neutrinos produced by the pion decay from the CERN-CNGS beam.

In a measurement  by MINOS \cite{Minos} it was found that muon neutrinos of average energy
3 GeV traversing a distance 730 km exceed $c$ by an amount, $\delta (E=3 GeV)= (5.1 \pm 2.9)
\times 10^{-5}$.  This however is in contrast to the neutrino observations from supernova
SN 1987a \cite{SN1,SN2} where over a flight path of $51$ kpc, the neutrinos with energy in
the band $(7.5 -39)$ MeV all arrived within a time span of $12.4$ sec and the optical
signal arrived after $4$ hours of the neutrino signal (consistent with prediction of
supernova models) from which it is inferred that $\delta (E=15 MeV) \leq 10^{-9}$.  

  In the present paper we calculate the rates for $e^+ e^-$ radiation from neutrinos
and pion decay assuming that the neutrinos obey the energy momentum relation of the form
$E^2=m^2 + p^2 + \xi_n p^n$. This is motivated by Horava-Lifshitz type field theories
\cite{Horava,Visser} where the higher derivative terms break Lorentz invariance at high
scale but help in removing ultraviolet divergence.
We find that for $n>2$ the bounds on $\xi_n$ from direct time of flight measurement \cite{Icarus12} are 
much more stringent than from constraining the Cerenkov decay process during the Cern to Gran-Sasso flight of
neutrinos \cite{Icarus}.  This is unlike the case for $n=2$ studied by Glashow-Cohen \cite{GC} and
\cite{Mestres, Bi, Cowsik,Altschul} where the constraint on Lorentz violation from the kinematically forbidden
processes is more stringent than from the time of flight measurement.

Models which explain the Opera result of superluminal neutrinos and which have a bearing
on the question of Cerenkov emission from neutrinos or the pion decay kinematics fall
broadly in the following  categories:

\begin{enumerate}

\item  Deformed Lorentz symmetry models \cite{Magueijo, Lingli, Amelino, Ling, Huo, Chang,
	Guo} where the dispersion relations change from the usual $p^2=m^2$ form to a
	different form which is still covariant under the modified Lorentz
	transformations.  In this picture the processes which are forbidden in one
	reference frame (like the rest frame of the massive neutrinos) will be forbidden
	also in the lab frame.

\item  Lorentz invariance violation as in Lifshitz type field theories
	\cite{Horava,Visser,Alexandre,Ellis-Alexandre, Saridakis}, from a gauge singlet
	SUSY sector \cite{Giudice} or a hidden sector\cite{Schreck}, environmental
	couplings \cite{Dvali,Kehagias, Matone, Mann, Oda, Sahu, Hebecker}, dynamical
	symmetry breaking \cite{Volovik,Nojiri}, Fermi-point splitting \cite{FPS},
	space-time fluctuations \cite{ellis2} and string theory \cite{Li}.

\end{enumerate}

In this paper we we consider the following general dispersion relation motivated by
Horava-Lifshitz theories
\begin{equation}
	E^2=m^2+p^2+\xi_n p^n
	\label{disp}
\end{equation}
where $n=2,3,4..$ etc.  The difference between the superluminal neutrino
velocity and the speed of light (taken to be 1) is then given by
\begin{equation}
	\delta=\frac{\partial E}{\partial p}-1 \simeq \frac{n-1}{2} \xi_n p^{n-2}, \quad n=2,3,4\dots
	\label{delta}
\end{equation}
The ICARUS time of flight experiment \cite{Icarus12}  has observed neutrinos and the time
difference between the neutrino time of flight (tof) and the calculated photon tof is $\delta t =0.3 \pm 4.9 (stat)\pm 9.3 (stat) $ 
for neutrino energy $E_\nu=12.5$ GeV and a distance 
$(731278.0\pm 0.2)m$ from CERN-CNGS to the detector in Gran Sasso.
This corresponds to a neutrino superluminality by the amount \cite{Opera},
\be
\delta(E=12.5{\rm GeV})= \frac{v_\nu-c}{c} < 2.3 \times 10^{-7}
\label{d17}
\ee
From (\ref{delta}) and (\ref{d17}) we put constraints on $\xi_n$ from the tof experiment \cite{Icarus12}.

The Cerenkov decay constraint comes from the earlier ICARUS experiment \cite{Icarus} where the expected number of CC neutrino events is $315 \pm 5$ and the observed number is
$308$. This corresponds to a bound on the decay length $c\tau$ of neutrinos given by
\be
 0.04 < \exp(- c \tau/731.2 {\rm km}). 
\label{cerenkov}
\ee
We calculate the neutrino Cerenkov  decay   
length $c \tau$ using the generalised dispersion relations (\ref{disp}) and constraint $\xi_n$ from the experimental bound (\ref{cerenkov}). 

The upper bounds on the Lorentz violating parameter 
 $\xi_n$ for different $n$ obtained from the tof experiment \cite{Icarus12} and Cerenkov decay experiment \cite{Icarus} are displayed in 
Table \ref{xin}. 
\begin{table}[ht]
	\centering
	\begin{tabular}{|c|c|c|}
	\hline
	\multirow{2}{*}{$n$} & \multicolumn{2}{|c|}{Upper bound on $\xi_n $} \\\cline{2-3} 
	 & Time of flight \cite{Icarus12} & Cerenkov process\cite{Icarus}\\\hline\hline
	2 & $4.6\times 10^{-7}$ & $1.7\times 10^{-5}$\\\hline
	3 & $5.9\times 10^{-8}$ GeV$^{-1}$ & $1.3\times 10^{-6}$ GeV$^{-1}$\\\hline
	4 & $3.8\times 10^{-10}$ GeV$^{-2}$ & $1.0\times 10^{-7}$ GeV$^{-2}$\\\hline
	5 & $2.5\times 10^{-11}$ GeV$^{-3}$ & $7.9\times 10^{-9}$ GeV$^{-3}$\\\hline
	6 & $1.7\times 10^{-12}$ GeV$^{-4}$ & $6.3\times 10^{-10}$ GeV$^{-4}$\\\hline
   	7 & $1.2\times 10^{-13}$ GeV$^{-5}$ & $5.0\times 10^{-11}$ GeV$^{-5}$\\\hline
	\end{tabular}
	\caption{  Upper bounds on the Lorentz violating parameter 
 $\xi_n$ for different $n$ obtained from the tof experiment \cite{Icarus12} and neutrino Cerenkov decay experiment \cite{Icarus}  }
	\label{xin}
\end{table}

\section{$\nu_\mu\to \nu_\mu e^+e^-$}

We compute the process $\nu_\mu(p)\to\nu_\mu(\p1)e^+(k)e^-(k^\prime)$ for GeV energy
neutrinos. We will use the dispersion relations
(\ref{disp}) for neutrinos in the Lab frame, and we will assume that all other particles
have the standard energy-momentum relations, and we will assume energy momentum
conservation in all reference frames. This generalises the Glashow-Cohen calculation for
the same process for the $n=2$ energy independent $\delta$ case.

The amplitude squared for the process is given by
\begin{align}
	\overline{|M|^2}=&32 G_F^2\left[(p\cdot k^\prime)(\p1\cdot k)\left(1-4\sw
	+8\sws\right)
	\right].
\label{modM}
\end{align}
The decay rate of the neutrino is in general given by
\begin{equation}
 \Gamma=\frac{1}{8(2\pi)^5}\int\:\frac{d^3\p1}{E_\nu^\prime}\,
	\frac{d^3k^\prime}{E_e^\prime}\,\frac{\overline{|M|^2}}{E_\nu}\,
	\delta\left((p-\p1-k^\prime)^2\right)
\label{rate1}
\end{equation}
Without loss of generality we can choose
\begin{align*}
 p&=(E_\nu,0,0,|{\bf p}|)\\
 \p1&=(E_\nu^\prime,|{\bf \p1}|\sin\theta,0,|{\bf \p1}|\cos\theta)\\
 k^\prime&=E_e^\prime(1,\cos\phi\sin\theta_1,\sin\phi\sin\theta_1,\cos\theta_1).
\end{align*}
The argument of the $\delta$ function in eq.(\ref{rate1}) can be  as
\begin{align}\nonumber
	(p-\p1-k^\prime)^2=\xi_n\left(p^{n-1}-(\p1)^{n-1}\right)(p-\p1)-p\p1\theta^2-
	\E1 D
\end{align}
where
\begin{equation}
 D=\xi_n\left(p^{n-1}-(\p1)^{n-1}\right)-\p1\left(\theta^2+\theta_1^2\right)
   +p\theta_1^2+2\p1\theta\theta_1\cos\phi
\end{equation}
Here we will assume that the transverse energy is very small and of the order of $\xi_n
p^{n-2}$, therefore we keep only the leading order terms in $\theta$, $\theta_1$ and
$\xi_n$.  From here on we shall use the notation $p$ and $\p1$ to denote $|{\bf p}|$ and
$|{\bf\p1}|$.

Now to fix the limits of the $\theta^2$ and $\theta_1^2$ integrals we need to find their
maximum values from the $\delta$-function condition i.e
\begin{equation}
 D\E1=\xi_n\left(p^{n-1}-(\p1)^{n-1}\right)(p-\p1)-p\p1\theta^2
\end{equation}
For the maximum value of $\theta$ we set $\E1=0$ in the above equation so that
have
\begin{equation}
	\theta_{max}^2=\frac{\xi_n\left(p^{n-1}-(\p1)^{n-1}\right)(p-\p1)}{p\p1}
\end{equation}

And similarly setting $p^\prime=0$ and the electron energy at its maximum i.e
$E_e^\prime=p/2$ in the $\delta$-function condition we have,
\begin{equation}
	(\theta_1^2)_{max}=\xi_n p^{n-2}\label{t1max}
\end{equation}

We make the following change of variables to pull out the factors of $\xi_n$ and $p$
from
the integrand :
\begin{equation*}
	\p1\to x\,p,\,\,\,,\theta\to\xi_n\,p^{n-2}\,\ta\,\,,\theta_1\to\xi_n\,p^{n-2}\,\tb
\end{equation*}
Using the above definitions in eq.(\ref{modM}) and substituting in eq.(\ref{rate1}) gives
the rate of electron-positron pair emission as
\begin{align}\nonumber
	\Gamma=\frac{G_F^2}{16\pi^4}\,\left(\xi_n p^{n-2}\right)^3 p^5&\left(1-4\sw +8\sws\right)
	\int_0^1\:dx\:\int_0^\frac{(1-x)(1-x^3)}{x}\:d\tilde{\theta}&\\
	&\int_0^1\:d\tilde{\theta_1}\,\int_0^{2\pi}\,d\phi\:
	f(x,\tilde{\theta},\tilde{\theta_1},\phi)&
\end{align}
where $f$ is complicated function of $x$, $\tilde{\theta}$ and $\tilde{\theta_1}$ \cite{mohanty}.

After numerically solving this integral we get the following expression for the rate of
electron-positron pair emission the general formula for the decay width of neutrino
splitting process $\nu(p)\to\nu(\p1)e^+(k)e^{-}(k^\prime)$ comes out to be
\be
	\Gamma=\frac{G_F^2}{16\pi^4 }\left(1-4\sw +8\sws\right)\,I_n\,
	\left[\xi_n\,p^{n-2}\right]^3 p^5
	\label{ee}
\ee
In eq.(\ref{ee}) $I_n$ is an integral of a function depending on $n$.  The values of
$I_n$ and $\Gamma$ for different values of $n$ are given in Table \ref{eepl}. Using $c \tau$ for different $n$ and comparing with the experimental bound from 
ICARUS\cite{Icarus} shown in (\ref{cerenkov}) we obtain the bounds on the Lorentz violating parameter $\xi_n$ for different $n$ displayed in Table \ref{xin}.


\begin{table}[ht]
	\centering
	\begin{tabular}{|c|c|c|c|c|c|c|c|}
	\hline
	$n$ & 2 & 3 & 4 & 5 & 6 & 7 & 8\\\hline\hline
	$I_n$ & $1/40$ & $1/31$ & $1/29$ & $1/28$ & $1/28$ & $1/28$ & $1/28$ \\\hline
	\end{tabular}
	\caption{Values of integral $I_n$ in eqn.(\ref{ee}) for different values of $n$ and for $E_\nu = 12.5$GeV.}
	\label{eepl}
\end{table}

This generalises  our earlier calculation of the $n=2$ and $n=4$ cases \cite{mohanty}.
Exact analytical calculations for the  $n=2$ case has been done \cite{GC, mohanty,
replaying, nano, Bezrukov}. Our result for the decay width is smaller than the
corresponding result of \cite{GC} and \cite{nano} by a factor of $2/3$ but is in closer
agreement with the results of \cite{replaying, Bezrukov}.

\section{Pion Decay}

We calculate the pion decay width in the lab frame  with a superluminal neutrino in the
final state. We assume the dispersion relation $E^2= (p^2+\xi_n p^n)$ in the lab
frame.   The amplitude squared for the process $\pi^-(q)\to\mu^-(p)\bar{\nu}_\mu(k)$ is,
\begin{equation}
 \overline{|M|^2}=4G_F^2f_\pi^2m_\mu^2\left[m_\pi^2-m_\mu^2+\xi_n k^n
 \left(\frac{m_\pi^2}{m_\mu^2}+2\right)\right]
\end{equation}

The decay width is then given by
\begin{align}
 \Gamma=&\frac{G_F^2f_\pi^2m_\mu^2}{8\pi E_\pi}\int\,
	\frac{k\,dk\,d\cos\theta}{\sqrt{|\vec{q}-\vec{k}|^2+m_\mu^2}}
	\delta(E_\nu+\sqrt{|\vec{q}-\vec{k}|^2+m_\mu^2}-E_\pi)\nonumber\\
	&\left[m_\pi^2-m_\mu^2+\xi_n k^n\left(\frac{m_\pi^2}{m_\mu^2}+2\right)\right]
\label{pi-rate}
\end{align}

\begin{table}
	\centering
	\begin{tabular}{|c|c|c|c|c|c|}
	\hline
	\multirow{2}{*}{$p_\pi$} &
	\multicolumn{5}{|c|}{$\Gamma/\Gamma_0$}\\\cline{2-6}
	  & $n=2$ & $n=3$ & $n=4$ & $n=5$ & $n=6$\\\hline\hline
	$10$ GeV & $0.95$ & $1.0$ & $1.0$ & $1.0$ & $1.0$ \\\hline
	$50$ GeV & $0.16$ & $0.48$ & $0.62$ & $0.69$ & $0.73$ \\\hline
	$100$ GeV & $0.04$ & $0.18$ & $0.27$ & $0.32$ & $0.35$ \\\hline
	$200$ GeV & $0.01$ & $0.06$ & $0.11$ & $0.13$ & $0.15$ \\\hline
	$500$ GeV & $0.002$ & $0.02$ & $0.03$ & $0.04$ & $0.05$ \\\hline
	\end{tabular}
	\caption{Ratio of pion decay width in Lorentz violating framework to its SM
	prediction ($\Gamma_0$) at different values of pion momentum ($p_\pi$) for various
	$n$.}
	\label{pitab}
\end{table}

Writing $|\vec{q}-\vec{k}|^2=k^2+q^2-2kq\cos\theta$, where $\theta$ is the angle between
$\vec{k}$ and $\vec{q}$, and $E_\nu=k+\xi_n k^{n-1}/2$ we see from the argument of the
$\delta$-function in eq.(\ref{pi-rate})
\begin{equation}
\cos\theta=\left(m_\mu^2-m_\pi^2+2E_\pi k+\xi_n k^{n-1}E_\pi-\xi_n k^n\right)(2kq)^{-1}
\label{cos}
\end{equation}
while the derivative of the argument of $\delta$-function with respect to $\cos\theta$
yields
\begin{equation}
\left|\frac{d}{d\cos\theta}(E_\nu+\sqrt{|\vec{q}-\vec{k}|^2+m_\mu^2}-E_\pi)\right|=
\frac{kq}{\sqrt{k^2+q^2-2kq\cos\theta+m_\mu^2}}
\end{equation}
Substituting this in eq.(\ref{pi-rate}) we get
\begin{equation}
 \Gamma=\frac{G_F^2f_\pi^2m_\mu^2}{8\pi E_\pi}\int\,\frac{dk}{q}
        \left[m_\pi^2-m_\mu^2+\xi_n k^n\left(\frac{m_\pi^2}{m_\mu^2}+2\right)\right]
\end{equation}

\begin{figure}[h!]
	\begin{center}
		\includegraphics[width=0.8\textwidth,clip=true]{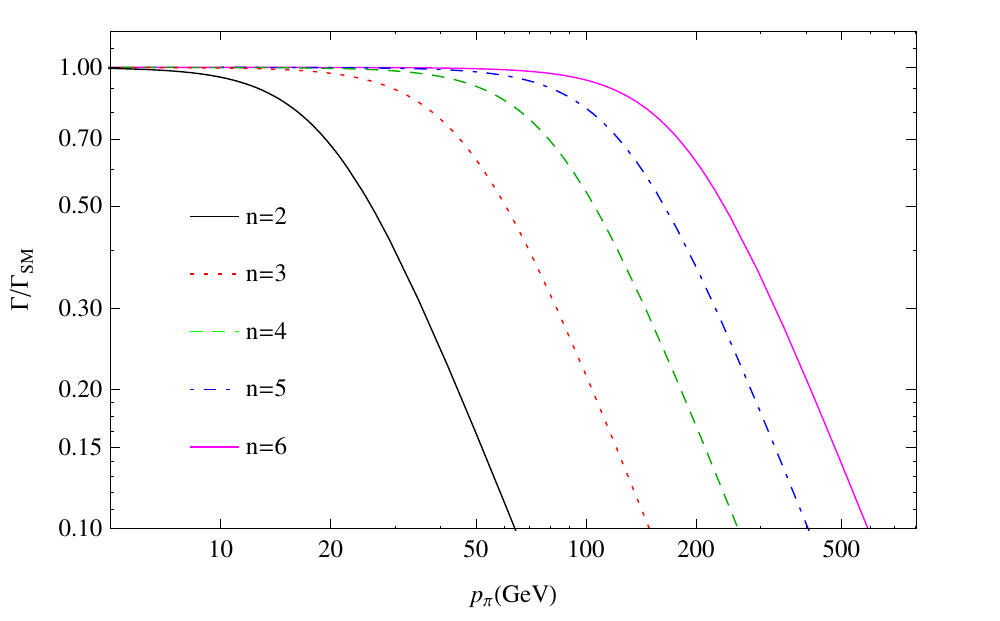}
	\end{center}
	\caption{Ratio of pion decay width in Lorentz violating framework (approximate
	numerical calculation) to its Standard
	Model prediction as a function of pion momentum ($p_\pi$) for different $n$.}
	\label{pifig}
\end{figure}

The limits of the $k$ integral are fixed by taking $\cos\theta=\pm 1$ in eq.(\ref{cos})
\begin{align}\nonumber
	k_{max}=&\frac{m_\pi^2-m_\mu^2-\xi_n k_{max}^{n-1}(E_\pi-k_{max})}{2(E_\pi-q)}\\[8pt]
 	k_{min}=&\frac{m_\pi^2-m_\mu^2-\xi_nk_{min}^{n-1}(E_\pi-k_{min})}{2(E_\pi+q)}
 \label{kmax}
\end{align}
we solve these polynomial equations for $k_{max}$ and $k_{min}$ numerically to obtain the
kinematically allowed limits of neutrino momentum.  Using these limits to integrate over
the neutrino momentum $k$ we get the decay rate for pion.  The ratio of pion decay rate
thus calculated to the standard model prediction
\be
\Gamma_0(\pi\to \mu \nu)=\frac{m_\pi^2}{E_\pi}\left(1-\frac{m_\mu^2}{m_\pi^2}\right)^2
\ee
for different $n$ is shown in the Table~\ref{pitab}.  The $n=4$ case has also been dealt
with in  \cite{Mannarelli:2011tv} and we are in broad agreement with their result. In
Fig.~\ref{pifig} an approximate numerical calculation of the pion decay width is plotted
as a function of pion momentum for different $n$.

The decay width for the 100 GeV pions 
decreases by $65\%$ for $n=6$ and is smaller at higher $n$.

\section{Conclusions}

We have computed neutrino processes assuming a power law correction to the neutrino
energy-momentum relations.  We conclude that for steep power
law ( $n\geq 2$) dispersion relations the contraint from the time of flight experiments is 
more stringent than from the measuremnet of the Cerenkov $e^+ e^-$ emission or from the change
in the width of pion decay.  Our calculation of neutrino Cerenkov emission and pion decay width
in Lorentz violating theories can be applied for putting bounds on Lorentz violating parameters
from the analysis of high energy cosmic rays.  Also, future experiments for measuring neutrino velocities
performed at higher energies will put strong constraints on the higher derivative Lorentz violation
theories \cite{Kostelecky}.

\end{document}